\documentclass[prb,preprint,showpacs,amssymb,amsmath,superscriptaddress,endfloats]{revtex4-1}

\usepackage{todonotes}
\usepackage{graphicx}
\usepackage{epstopdf}
\usepackage{color}
\usepackage{changes}
\usepackage{hyperref}

\definecolor{red}{rgb}{1.00,0.00,0.00}
\graphicspath{ {./figures/} }

\newcommand{\mvec}[1]{\ensuremath{\mathbf{#1}}}

\usepackage{listings}

\definecolor{dkgreen}{rgb}{0,0.6,0}
\definecolor{gray}{rgb}{0.5,0.5,0.5}
\definecolor{mauve}{rgb}{0.58,0,0.82}

\begin{document}

\title{Proposal of a micromagnetic standard problem for ferromagnetic resonance simulations}

\author{Alexander Baker} %\email{alexander.baker@physics.ox.ac.uk}
\affiliation{Department of Physics, Clarendon Laboratory, University of Oxford, Oxford OX1~3PU, United Kingdom}
\author{Marijan Beg} % \email{mb4e10@soton.ac.uk}
\author{Gregory Ashton} % \email{gash789@gmail.com}
\author{Maximilian Albert} % \email{maximilian.albert@gmail.com}
\author{Dmitri Chernyshenko} % \email{d.chernyshenko@soton.ac.uk}
\affiliation{Faculty of Engineering and the Environment, University of Southampton, SO17~1BJ, Southampton, United Kingdom}
\author{Weiwei Wang} % \email{ww1g11@soton.ac.uk}
\affiliation{Department of Physics, Ningbo University, Ningbo 315211, China}
\author{Shilei Zhang} % \email{shilei.zhang@physics.ox.ac.uk}
\affiliation{Department of Physics, Clarendon Laboratory, University of Oxford, Oxford OX1~3PU, United Kingdom}
\author{Marc-Antonio Bisotti} % \email{mb8g11@soton.ac.uk}
\author{Matteo Franchin} % \email{franchin@soton.ac.uk}
\affiliation{Faculty of Engineering and the Environment, University of Southampton, SO17~1BJ, Southampton, United Kingdom}
\author{Chun Lian Hu}
\author{Robert Stamps} %\email{robert.stamps@glasgow.ac.uk}
\affiliation{SUPA School of Physics and Astronomy, University of Glasgow, Glasgow G12~8QQ, United Kingdom}
\author{Thorsten Hesjedal} %\email{t.hesjedal1@physics.ox.ac.uk}
\affiliation{Department of Physics, Clarendon Laboratory, University of Oxford, Oxford OX1~3PU, United Kingdom}
\author{Hans Fangohr} \email{fangohr@soton.ac.uk}
\affiliation{Faculty of Engineering and the Environment, University of Southampton, SO17~1BJ, Southampton, United Kingdom}

\begin{abstract}
Nowadays, micromagnetic simulations are a common tool for studying a wide range of different magnetic phenomena, including the ferromagnetic resonance. A technique for evaluating reliability and validity of different micromagnetic simulation tools is the simulation of proposed standard problems. We propose a new standard problem by providing a detailed specification and analysis of a sufficiently simple problem. By analyzing the magnetization dynamics in a thin permalloy square sample, triggered by a well defined excitation, we obtain the ferromagnetic resonance spectrum and identify the resonance modes via Fourier transform. Simulations are performed using both finite difference and finite element numerical methods, with \textsf{OOMMF} and \textsf{Nmag} simulators, respectively. We report the effects of initial conditions and simulation parameters on the character of the observed resonance modes for this standard problem. We provide detailed instructions and code to assist in using the results for evaluation of new simulator tools, and to help with numerical calculation of ferromagnetic resonance spectra and modes in general.
\end{abstract}


% 75.40.Mg    Numerical simulation studies
% 76.50.+g    Ferromagnetic, antiferromagnetic, and ferrimagnetic resonances; spin-wave resonance
% 75.70.-i    Magnetic properties of thin films, surfaces, and interfaces

\maketitle

%Table of contents for better understanding of paper structure
%\tableofcontents

\section{Introduction\label{sec:introduction}}

Computational micromagnetics is a well developed field that sees widespread use in both modern physics and magnetic device engineering communities.\cite{Berkov:MicromagExampls, Erokhin:MicromagExample, Finocchio:MicromagExample} With the advancement of micromagnetic models, simulation techniques, and processing power, the list of phenomena that can be studied has grown substantially and includes such diverse fields as the spin transfer torque\cite{Najafi:spinTorque} and spin wave dispersion in magnonic crystals.\cite{Ma:magnonic} An essential equation in most of the micromagnetic system models\cite{Brown:MMS} is the Landau-Lifshitz-Gilbert (LLG) equation -- a differential equation governing the magnetization dynamics. However, this equation can be analytically solved only for a very limited number of systems and, because of that, the complexity of common problems requires the use of micromagnetic simulation packages such as \textsf{OOMMF}\cite{Donahue:OOMMF}, \textsf{LLG Micromagnetics},\cite{Scheinfein:LLG} \textsf{Micromagnum},\cite{micromagnum} and \textsf{Mumax}\cite{vansteenkiste:muMax}, which use the Finite Difference (FD) approach, and \textsf{Nmag}\cite{Fischbacher:Nmag} and \textsf{Magpar},\cite{Scholz:Magpar} employing the Finite Element (FE) approach to spatial discretization. To compare this range of numerical solvers, as well as to evaluate their validity and reliability, NIST's Micromagnetic Modelling Activity Group ($\mu$Mag) publishes standard problems.\cite{Donahue:muMag2,Hertel:muMag3,Tsiantos:muMag4} Recent additions have included the spin transfer torque \cite{Najafi:spinTorque} and the spin wave dispersion\cite{Venkat:spinWaveDispersal} standard problems. In the light of this, it is natural to extend the coverage of standard problems in order to include the FerroMagnetic Resonance (FMR), a technique closely associated with many practical uses ranging from material characterization to the study of spin dynamics.\cite{farle_FMRReview}

FMR probes the magnetization dynamics in samples using microwave fields. The absorption of the applied microwave field is at its maximum when the microwave's frequency matches the frequency of the studied system's resonant modes. By analyzing the resonance modes as a function of an applied magnetic field, some material parameters, such as the Gilbert damping and magnetic anisotropy constants, can be determined.\cite{farle_FMRReview} This makes FMR a powerful technique in the characterization of ferromagnetic nanostructures; including measurements of spin pumping\cite{heinrich_spinPumping} and exchange coupling.\cite{heinrich_exchangeCoupling} In a typical experiment, microwaves are directed across the sample using a coplanar waveguide, and their transmission is measured as a function of both  external bias field and excitation frequency.\cite{Nembach:FMRExperiment}

In terms of computational micromagnetics, there are at least three methods that can be used to simulate the FMR:
\begin{enumerate}
\item Application of a time-dependent periodic sinusoidal magnetic microwave field of fixed frequency $f$ to determine the magnetization precession amplitude in response to the system. If the precession amplitude is small, the power absorption of the microwave field would be small as the excitation frequency does not couple well to the set of natural frequencies of the system. This method is conceptually simple but computationally very demanding as, for every frequency $f$, the micromagnetic simulation needs to compute the time evolution of the system's magnetization after the transient dynamics has been damped and steady magnetization precession is reached. This will only provide one point on the frequency-absorption curve and only a micromagnetic simulation software that supports a time dependent external magnetic field can be used.

\item Ringdown method:\cite{McMichael2004} the system is perturbed from its equilibrium state by applying a short-lived and sufficiently weak excitation, followed by simulation and recording of the magnetization dynamics. Resonance frequencies and corresponding modes are extracted by performing the Fourier transform on the recorded data. This is an efficient way to determine the eigenmodes of the system.

\item Eigenvalue method:\cite{dAquino:NormalModes} instead of simulating the time evolution of the system's magnetization as in the methods above, the problem is represented as an eigenvalue problem, whose solutions provide the frequencies (eigenvalues) and mode shapes (eigenvectors) of the system. This method requires specialist software that is not widely available.
\end{enumerate}

Our goal is to establish a standard problem to serve as a benchmark against which future simulation tools and computational studies of the FMR can be compared and validated. In this standard problem proposal, we will follow the second (ringdown) method, which is supported by most micromagnetic packages and compare its output with the third (eigenvalue) method. We provide a detailed standard problem description and specification as well as the complete set of computational steps and code repository\cite{Githubrepo2015} in order to make it easily reproducible and accessible to a wide community. Parts of the code repository can also be used as an example to compute FMR data and modes from micromagnetic simulations. It is hoped that this work will aid the development of micromagnetic simulations of systems undergoing FMR and support and drive experimental efforts.

\medskip
Sec.~\ref{sec:problem} introduces and motivates the choice of the FMR standard problem, and introduces the frequency spectrum computed in different ways. Sec.~\ref{sec:discussion} provides a more detailed discussion including computation of the normal mode shape, the eigenvalue problem approach as an alternative way of computing the frequency spectrum and normal modes, and a systematic study of the dependence of the results on variations in the simulation parameters such as damping, relaxation of the initial state, nature of the perturbation and mesh discretization. We close with a summary in Sec.~\ref{sec:summary}. The Appendix provides more details on parameters used in the \textsf{Nmag} simulations, the eigenvalue approach and simulation results obtained in the absence of demagnetization effects.

\section{Selection and definition of standard problem\label{sec:problem}}

\subsection{Problem definition\label{subsec:problemdef}}

\begin{figure}
  \includegraphics{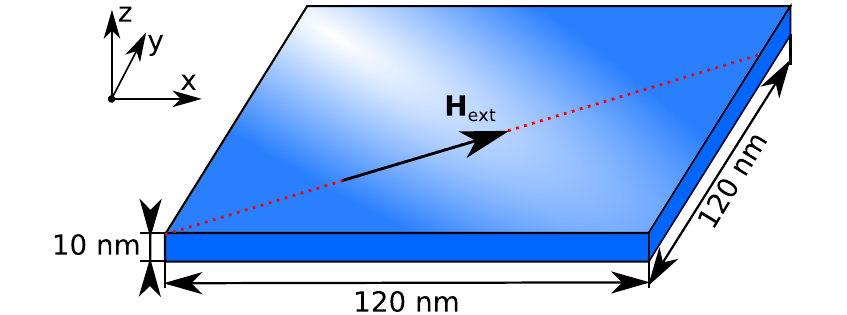}
  \caption{\label{fig:sample_geometry} Geometry of the thin film sample, showing the static bias field $\mvec{H}_\text{ext}$. The field is slightly off-diagonal to break the symmetry of the system and thus avoid degenerate eigenmodes.}
\end{figure}

We choose a cuboidal thin film permalloy  sample measuring $120 \times 120 \times 10 \,\text{nm}^{3}$, as shown in Fig.~\ref{fig:sample_geometry}. The choice of a cuboid is important as it ensures that the finite difference method employed by \textsf{OOMMF} does not introduce errors due to irregular boundaries that cannot be discretized well.\cite{Donahue2007} We choose the thin film geometry to be thin enough so that the variation of magnetization dynamics along the out-of-film direction can be neglected. Material parameters based on permalloy are shown in Table~\ref{tab:material_parameters}. An external magnetic bias field $\mvec{H}_\mathrm{ext}$ with magnitude $H_\text{ext} = 80 \,\text{kA/m}$ is applied along the direction $\mvec{e} = (1, 0.715, 0)$ (at $35.56^{\circ}$ to the $x$-axis), i.e. $\mvec{H}_\mathrm{ext} = H_\text{ext} \cdot \mvec{e}/|\mvec{e}| \approx (65.1, 46.5, 0) \,\text{kA/m}$ as shown in Fig.~\ref{fig:sample_geometry}. We choose the external magnetic field direction slightly off the sample diagonal in order to break the system's symmetry and thus avoid degenerate eigenmodes.

\begin{table}
  \begin{tabular}{l r l }
    \hline
    \hline
    Parameter & Value & Unit \\
    \hline
    saturation magnetization ($M_\text{s}$) &  800 & kA/m \\
    exchange constant ($A$) &  $1.3 \times 10^{-11}$ & J/m \\
    anisotropy constant ($K$) &  0 & J/m$^{3}$ \\
    gyromagnetic ratio ($\gamma^{*}$) &  $2.210173 \times 10^{5}$ & m/(As) \\
    Gilbert damping ($\alpha$), relaxation &  $1.0$ &\\
    Gilbert damping ($\alpha$), dynamic &  $0.008$ &  \\
    DC bias field magnitude ($|\mvec{H}_{0}|$) & 80 & kA/m \\
    DC bias field ($\mvec{e}$), relaxation &  [1, 0.715, 0]& \\
    DC bias field ($\mvec{e}$), dynamic & [1, 0.7, 0] & \\
    \hline
    \hline
  \end{tabular}%
  \caption{\label{tab:material_parameters} External magnetic fields and material (permalloy) parameters used. Where these change between the initial relaxation stage of the simulation, and the subsequent dynamic stage, both values are shown.}
\end{table}

First, we initialize the system with a uniform out-of-plane magnetization $\mvec{m}_{0} = (0, 0, 1)$. The system is allowed to relax for $5 \,\text{ns}$, which was found to be sufficient time to obtain a well-converged equilibrium magnetization configuration. We refer to this stage of simulation as the \textit{relaxation stage}, and its final relaxed magnetization configuration is saved to serve as the initial configuration for the next \textit{dynamic stage}. Conceptually, what is required to find the relaxed state is to minimize the system's energy in the presence of an external magnetic bias field, taking into account exchange and demagnetization energy contributions. We note that there are other ways of obtaining this configuration, including energy minimization (as for example supported by \textsf{OOMMF}), or solution of the LLG without the precession term (as supported by \textsf{Nmag}). Because we want to use a well defined method that is supported by all simulation tools, we minimize the system's energy by integrating the LLG equation with a large, quasistatic Gilbert damping $\alpha =1$ for $5 \,\text{ns}$. The use of any of these methods is expected to lead to the same relaxed equilibrium magnetization configuration.

In the next step (\textit{dynamic stage}), a simulation is started using the equilibrium magnetization configuration from the relaxation stage as the initial configuration. Now, the direction of an external magnetic field is altered to $\mvec{e} = (1, 0.7, 0)$, i.e. $\mvec{H}_\mathrm{ext} = H_\text{ext}(\mvec{e}/|\mvec{e}|) \approx (65.5, 45.9, 0) \,\text{kA/m}$. This corresponds to a rotation of the bias field to $35^{\circ}$ with respect to the $x$-axis. Due to the change in $x$ and $y$ components of the external magnetic field, the initial magnetization configuration is now out of equilibrium. Consequently, the system tends to relax towards the lowest energy configuration in the presence of a new external magnetic field. This simulation stage runs for $T = 20 \,\text{ns}$ while the (average and spatially resolved) magnetization $M(t)$ is recorded every $\Delta t = 5 \,\text{ps}$. The Gilbert damping in this dynamic simulation stage is $\alpha = 0.008$. Using the recorded data, a Fourier transform is performed to produce the FMR spectrum and obtain eigenfrequencies (and the eigenmodes). Spatially resolved transformations allow examination of the shapes of the modes (see Sec.~\ref{subsec:data}). Simulation parameters for both stages of the simulation are given in Tab.~\ref{tab:material_parameters}.

\subsection{Problem Selection\label{subsec:problemselect}}

In this section, we address the selection criteria for the standard problem, and give an explanation of how each is met within the proposed framework:

\begin{enumerate}
\item \textit{Initial magnetization configuration}. This standard problem is defined in two stages: (i) relaxation stage and (ii) dynamic stage. The purpose of the relaxation stage is to bring the system into a well defined state. Starting from an initial uniform out-of-plane magnetization $\mathbf{m}_0 = (0,0,1)$ combined with the in-place bias field $\mathbf{H}_0$, the system transitions into a ``relaxed'' state in an attempt to reach a (local) energy minimum. The relaxed state is used as the initial configuration for the dynamic stage.

\item \textit{Excitation of system}. Apart from being reproducible, the perturbation or excitation field must be sufficiently large to excite magnetization dynamics, yet be small enough so that the system remains in the linear regime. This is achieved by altering the direction of the bias field, as a simple practical approach that does not require time-dependent applied fields. The power spectrum obtained is specific for the chosen excitation, and thus the excitation is a key part of the problem definition. All simulations tools, even the ones that do not support time-dependent external magnetic fields, are expected to be able to excite the system in this manner.

\item \textit{Computation time}. Standard problems, apart from being simple and reproducible, require as short as possible computation time. In micromagnetic simulations, the computational time depends mostly on the number of degrees of freedom in the discretized problem. Accordingly, the spatial discretization of $5 \,\text{nm}$ is chosen as a balance between computational time and accuracy. Although the second simulation stage is performed with realistic Gilbert damping value $\alpha=0.008$ over a limited simulation time, in the first (relaxation) stage, we set $\alpha = 1$ to ensure the magnetization reaches a well converged state within the allotted time.

\item \textit{Verification of results}. Ideally, results should be verified against other methods of obtaining them. In this work, we use different simulation packages (including finite difference and finite element discretization schemes) that have been developed by different groups. Furthermore, we use a completely different computational (eigenvalue based) method to obtain the power density spectrum and excited normal modes separately.

\end{enumerate}

\subsection{Data Analysis\label{subsec:data}}

We outline two different ways to compute the power spectrum of the simulated system.

\subsubsection*{Method 1: Global power spectrum and $S_y(f)$\label{sec:four-transf-spat}}

In this case, the observable we use is the spatially averaged magnetization $\langle \mvec{M} \rangle_{\mvec{r}}(t)$, as it is easily accessible in all known simulation tools. Using a discrete Fourier transform,\cite{Press:NumericalRecipes} we can obtain the power spectrum of the average magnetization in the frequency domain. As the dynamic simulation progresses, at uniform time steps $t_{k}$, we record the spatially averaged magnetization $\langle \mvec{M} \rangle_{\mvec{r}}(t_{k})$, where $t_{k} = k \Delta t$ with $\Delta t = 5 \,\text{ps}$, and $k = 1, 2, \ldots, N$, with $N=4000$ being the number of time steps. However, we only consider the $y$-component of spatially averaged magnetization $\langle M_{y} \rangle_{\mvec{r}}(t_{k})$ to compute the power spectrum $S_y(f)$ using
\begin{eqnarray}
  S_y(f) &=& |\mathcal{F}_y(f)|^2 \quad \mathrm{with}\\
  \mathcal{F}_{y}(f) &=& \sum_{k=1}^{N} \langle M_{y} \rangle_{\mvec{r}}(t_{k}) e^{-i2\pi ft_{k}}.
    \label{Eq:spatiallyAveragedFT}
\end{eqnarray}
According to the chosen parameter values, the sampling frequency is $f_\text{s} = 1/\Delta t = 50 \,\text{MHz}$, which implies that the maximum frequency that can be sampled (Nyquist frequency\cite{Press:NumericalRecipes}) is $f_\text{N} = 2f_\text{s} = 100 \,\text{GHz}$. We term this approach ``method 1''. It requires that the discrete Fourier transform is performed once (on the time series of the average magnetization) in order to compute the power spectrum $S_y(f)$.

\subsubsection*{Method 2: Local power spectrum and $\tilde{S}_{y}(f)$}

Equation (\ref{Eq:spatiallyAveragedFT}) uses the spatially averaged magnetization to compute its frequency spectrum. Following McMichael and Stiles' approach\cite{McMichael2004} to compute a collection of local power spectra over the extent of the sample we introduce a second method which allows to gain more detailed information about the spectrum. In contrast to the first method, this requires computation of discrete Fourier transforms at all spatial sampling points.

We analyze $n = n_{x}n_{y}$ scalar time-dependent signals: for every recording time $t_k$ we sample the magnetization on a two-dimensional grid of positions $\mvec{r}_{m,p}$ where $n_{x}$ and $n_{y}$ are the number of sampling points in $x$ and $y$ directions, respectively. More precisely, $\mvec{r}_{m,p} = ((m-\frac{1}{2}) \frac{L_{x}}{n_{x}}, (p-\frac{1}{2}) \frac{L_{y}}{n_{y}} , 2.5\,\mathrm{nm})$ with $m = 1, 2, \ldots, n_{x}$, $p = 1, 2, \ldots, n_{y}$, and $L_{x} = L_{y} = 120\,\text{nm}$. In the remainder of this work, we have used $n_{x} = 24$ and $n_{y} = 24$. For simplicity and generality, we label the sampling points $\mvec{r}_{m,p}$ as $\mvec{r}_j$, with $j = 1, 2, \ldots, n_{x}n_{y}$.

\begin{figure}
  \includegraphics[width=1\columnwidth]{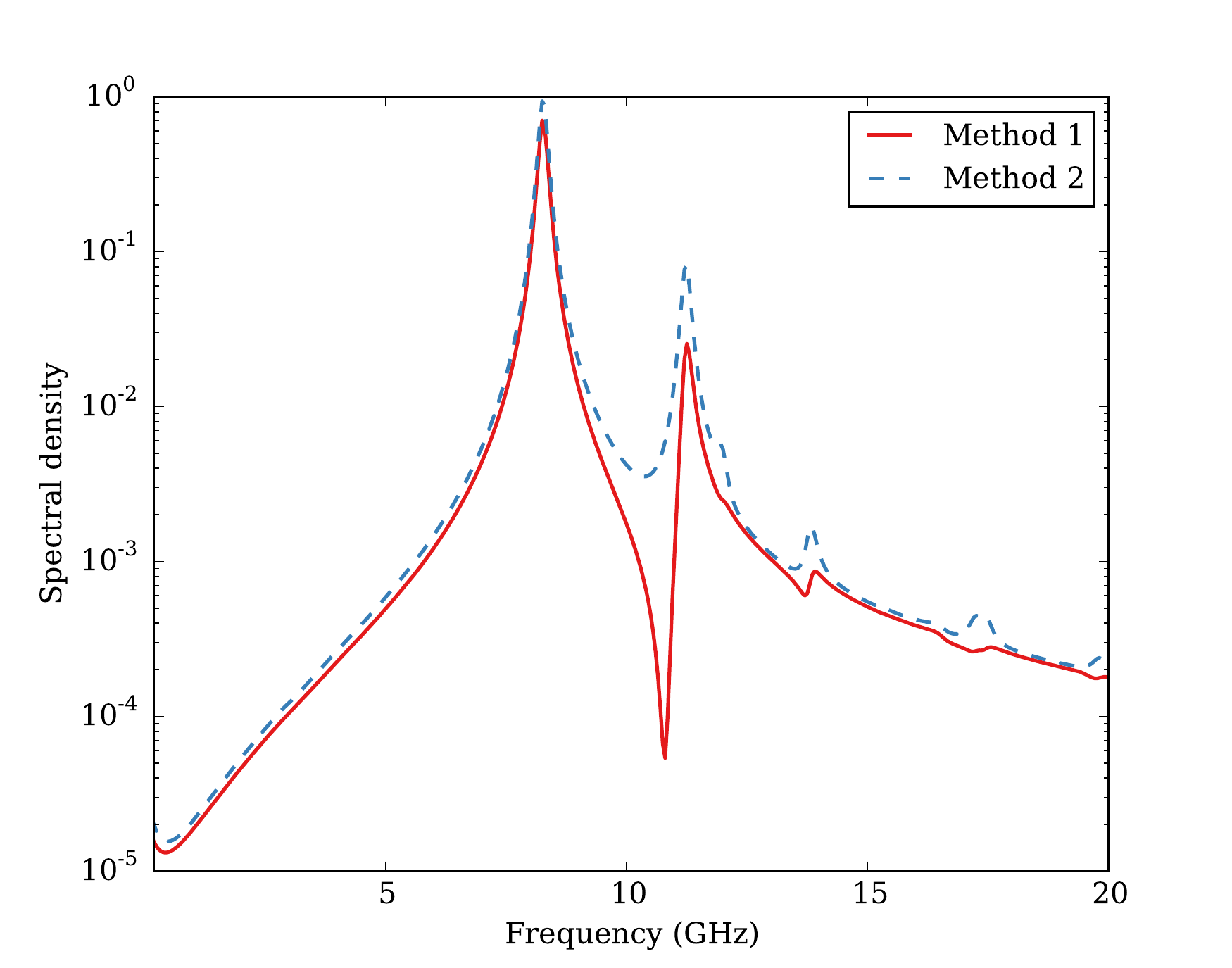}
  \caption{Power spectrum $S_{y}(f)$ from Eq.\ (\ref{Eq:spatiallyAveragedFT}) (method 1, solid red line) and $\tilde{S_{y}}(f)$ from Eq.\ (\ref{Eq:spatiallyResolvedFT}) (method 2, dashed blue line) from ringdown method in \textsf{OOMMF}.}
  \label{Fig:FTComparison}
\end{figure}

We term this approach ``method 2'', and compute the \emph{local} power spectrum
\begin{equation}
  S_{y}(\mvec{r}_{j},f) = |\mathcal{F}_{y}(\mvec{r}_{j},f)|^2
\end{equation} for each of the recorded signals (i.e. for each position $\mvec{r}_j$), with
\begin{equation}
  \mathcal{F}_{y}(\mvec{r}_{j},f)= \sum_{k=1}^{N}M_{y}(\mvec{r}_{j},t_{k})e^{-i2\pi ft_{k}}.
\label{Eq:localSpatiallyResolvedFT}
\end{equation}
By averaging the local power spectra $S_{y}(\mvec{r}_j, f)$, we obtain
\begin{equation}
  \label{Eq:spatiallyResolvedFT}
  \tilde{S}_{y}(f)=\frac{1}{n_{x} n_{y}} \sum_{j=1}^{n_{x} n_{y}} S_{y}(\mvec{r}_j, f).
\end{equation}
Both entities $S_{y}(f)$ and $\tilde{S}_{y}(f)$ are shown in Fig.~\ref{Fig:FTComparison} in a logarithmic scale, and strong resonance peaks are observed at $f_1 = 8.1 \,\text{GHz}$ and $f_2 = 11.1 \,\text{GHz}$.

This method allows us to obtain spatially resolved information $S_y(\mathbf{r}, f)$
about the normal modes of the system. See further discussion in
Sec.~\ref{subsec:standard} and Figs.~\ref{Fig:spatialModes_8} and
Fig.~\ref{Fig:spatialModes_11}.

\begin{figure}
  \begin{center}
    \includegraphics[trim = 10 40 10 10, width=1\columnwidth]{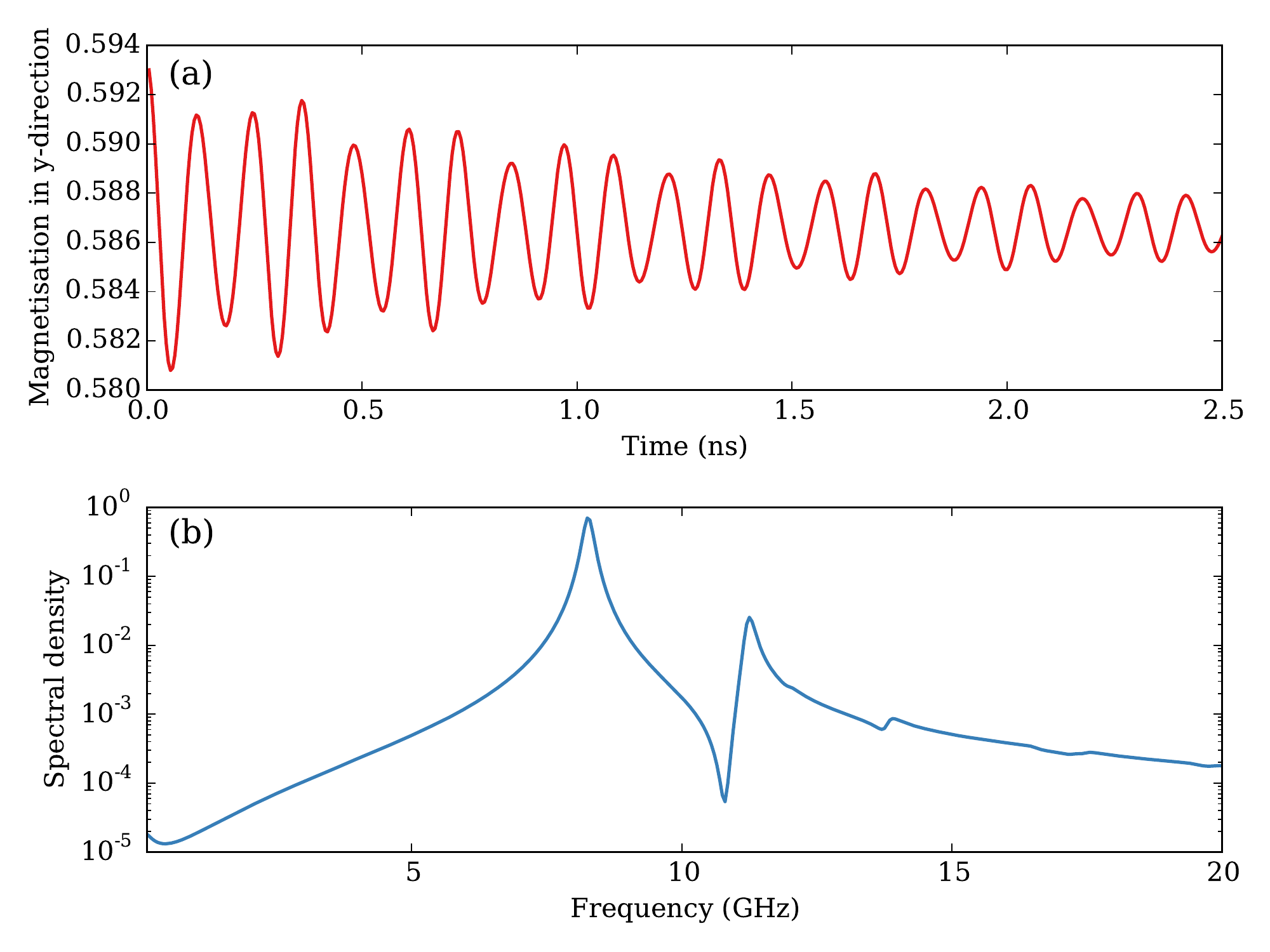}
  \end{center}
  \caption{\label{Fig:yMag} (a) Spatially averaged (method 1) $y$-component $\langle M_{y} \rangle_{\mvec{r}}(t)$ of the magnetization $M_{y}(t)$, as determined by the ringdown method in by \textsf{OOMMF}. (b) Power spectrum $S_y(f)$ obtained from Fourier transform of the spatially averaged y-component of the magnetization
  $\langle M_{y} \rangle_{\mvec{r}}(t)$ data, calculated using Eq.\ \ref{Eq:spatiallyAveragedFT}.}
\end{figure}

\subsubsection{Phase information\label{sec:phase-information}}

In order to understand the precession of a resonance mode $q$ at a particular frequency $f_q$ across the extent of a thin film, we need to extract the phase information from the spatially resolved Fourier transform. We start with the complex Fourier coefficient $\mathcal{F}_{y}(\mvec{r}_j, f_q)$ which represents the contribution of the frequency $f_q$ to the time series of the magnetization $y$-component $\mvec{M}_{y}(\mvec{r}_j, t)$ of the magnetization dynamics at position $\mvec{r}_j$. In our discrete Fourier transform, we have a set of $N$ complex Fourier coefficients $\mathcal{F}_{y}(\mvec{r}_j, f)$ at discrete frequencies $f_k$. The modulus (= absolute value) of the Fourier coefficient contains the information about the amplitude, whereas its argument (in the polar representation) contains the phase information. Consequently, the information about the resonance mode $q$ phase can be extracted as the complex Fourier coefficient argument as a function of position $\mvec{r}_j$, which allows us to identify the relative phases between different spatial domains in a normal mode.

\section{Results and discussion\label{sec:discussion}}

\subsection{Standard Problem Simulation Results\label{subsec:standard}}

Figures \ref{Fig:yMag} and \ref{Fig:FTComparison} show the main results from the standard problem, as outlined in Sec.~\ref{subsec:problemdef}, obtained using the \textsf{OOMMF} simulation tool. Time evolution of the average magnetization $y$-component for the first $2.5 \,\text{ns}$ of dynamic stage is shown in Fig.~\ref{Fig:yMag}(a), and the associated ferromagnetic resonance spectrum (Fourier transform of $\langle M_{y} \rangle_{\mvec{r}}(t)$ over the entire $20 \,\text{ns}$ dynamic simulation) is shown in Fig.~\ref{Fig:yMag}(b). Performing the Fourier transform of spatially averaged magnetization (method 1) produces a slightly different result in comparison to the spatially resolved (method 2) approach, which is shown in Fig.~\ref{Fig:FTComparison}.

\begin{figure}
  \includegraphics[width=1\columnwidth]{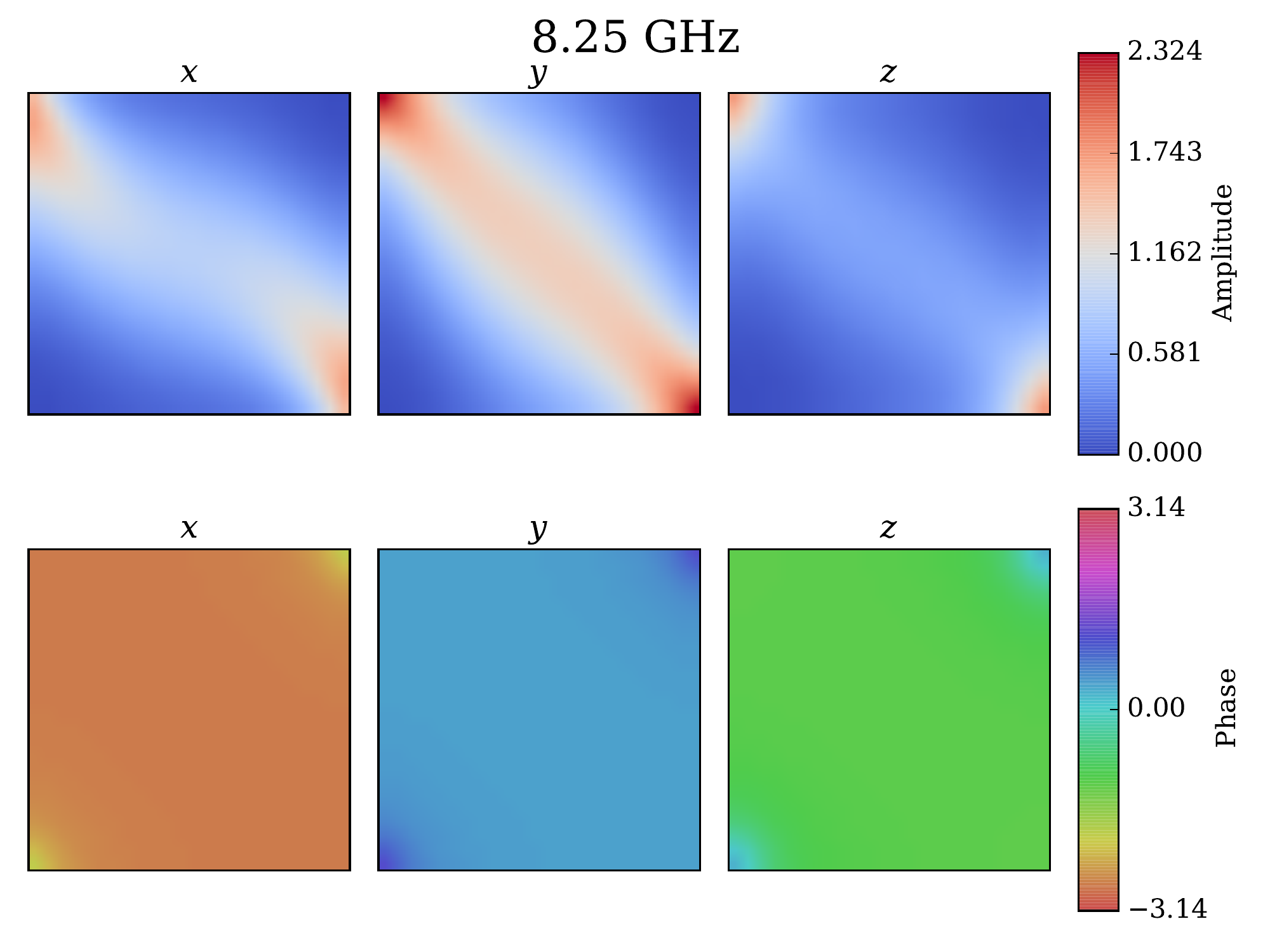}
  \caption{\label{Fig:spatialModes_8} Spatially resolved resonance modes in all three Cartesian directions plotted over the extent of the sample at $f_1 = 8.25\,$GHz obtained from ringdown method in \textsf{OOMMF}. Top row: base 10 logarithmic scale of power spectra for $x$-, $y$- and $z$-component, respectively.
Bottom row: corresponding phase distributions for three components.}
\end{figure}

Using the spatially resolved approach, one can plot the power spectrum coefficients $S_{y}(\mvec{r}_j, f_q)$ as a function of position $\mvec{r}_j$ for the normal mode frequency $f_q$ to represent both the power amplitude and phase of the normal mode $q$, as described in Sec.~\ref{sec:phase-information}. Figure~\ref{Fig:spatialModes_8} shows the spatial resolution of the resonance mode at $f_1 = 8.1 \,\text{GHz}$ with both the amplitude $|S|(\mvec{r}_j, f_1)$ and phase information $\arg(S)(\mvec{r}_j, f_1)$ for $x$, $y$ and $z$ magnetization components that were calculated from the \textsf{OOMMF} simulation using Eq.~(\ref{Eq:spatiallyResolvedFT}). The magnetization precession is present in all three directions, with the highest amplitude in the $y$-direction as expected since the largest external bias field perturbation is performed along the $y$-direction. Figure~\ref{Fig:FTComparison} shows that the frequency spectrum is dominated by two modes. The low frequency mode extends across the middle of the sample; this corresponds to the mode of uniform precession observed in macroscopic samples.

\begin{figure}
  \includegraphics[width=1\columnwidth]{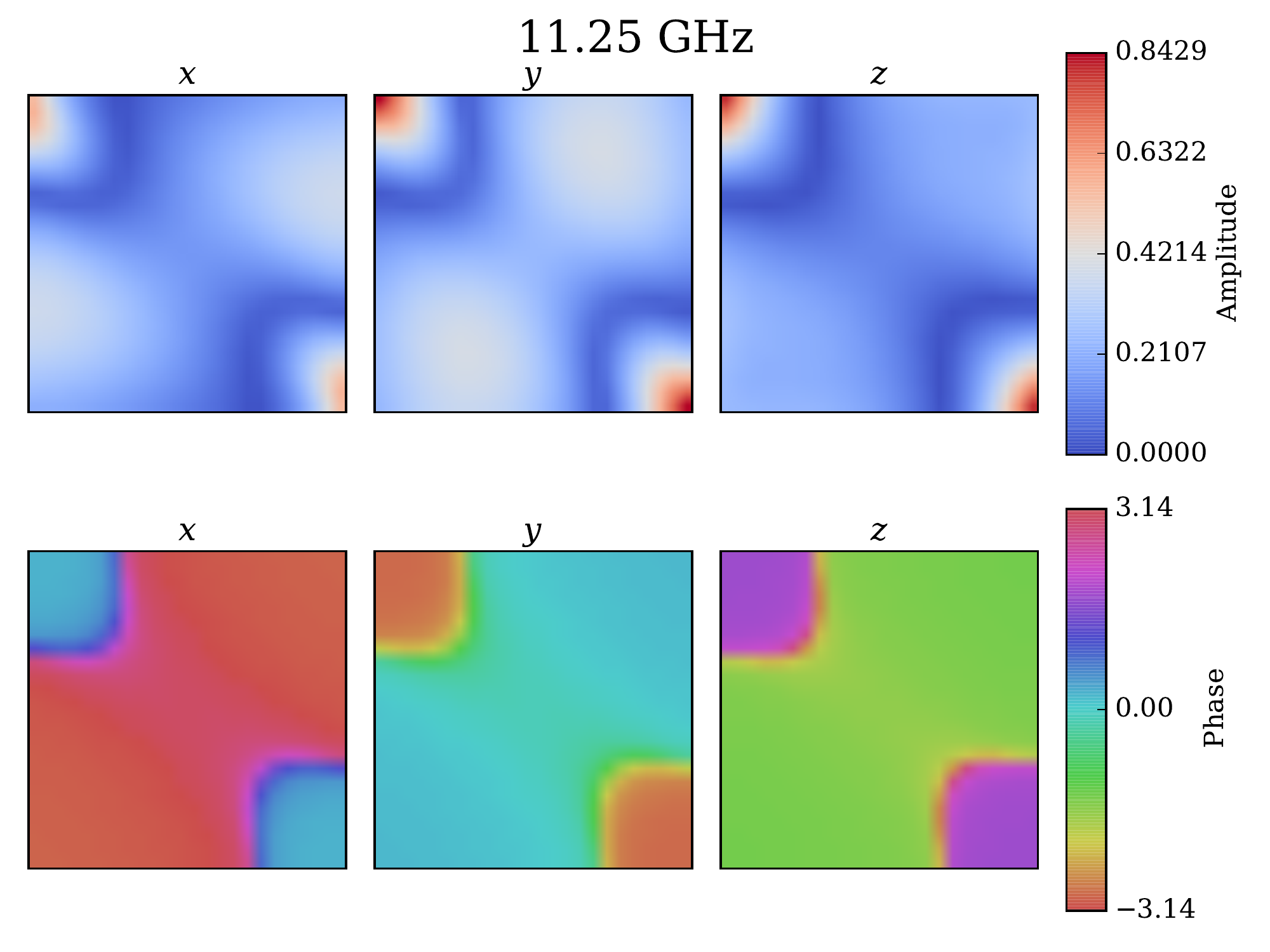}
  \caption{\label{Fig:spatialModes_11} Spatially resolved resonance modes in all three Cartesian directions plotted over the extent of the sample at $f_2 = 11.25\,$GHz obtained from ringdown method in \textsf{OOMMF}. Top row: base 10 logarithmic scale of power spectra for $x$-, $y$- and $z$-component, respectively.
Bottom row: corresponding phase distributions for three components.}
\end{figure}

The largest precession amplitude of the normal mode at $11 \,\text{GHz}$ (spatially resolved plot shown in Fig.~\ref{Fig:spatialModes_11}) is located at the corners of the sample and is dominated by the demagnetization energy associated with magnetization canting at the sample boundaries. In terms of the normal mode phase representation, an abrupt phase shift occurs as one moves away from the sample corner to the sample center. This normal mode is associated with the particular shape and size of the sample. Note that the precession amplitude in Fig.~\ref{Fig:spatialModes_8} (top row) and \ref{Fig:spatialModes_11} (top row) is generally small where the phase changes: these oscillation nodes separate domains that show out-of-phase precession relative to each other. Similar effects have been observed, for example, in permalloy nanodisks: Guo~\textit{et al.}\cite{Guo:edgeModes} used ferromagnetic resonance force microscopy to spatially resolve the resonance modes. They observed the same mode shapes simulated here, and demonstrated a strong relationship between the size of the disk and the relative strength of the modes. Appendix~\ref{subsec_appendix_noDemag} details the results of simulations performed without the demagnetization energy contribution (only one resonance is observed, corresponding to a macrospin model of uniform coherent precession).

A resonance mode also exists in the $z$-direction, $\mathcal{F}_{z}(\mvec{r}_{j},f)$. The precession of the moments describes an ellipse around the bias field, which has greatest amplitude in the $x-y$ plane, with the component in $z$ being smaller due to the demagnetization field.

\subsection{Eigenvalue method results \label{subSec:eigenvalueApproach}}

An alternative approach to calculating the normal modes is to linearize the LLG equation for the studied system around its equilibrium state; the normal modes of the resulting linear system of equations can then be determined by solving an eigenvalue problem. This approach does not require running and post-processing of a dynamic micromagnetic simulation, and is thus a good way to check the veracity of the results. A detailed description of this method providing resonance frequencies and normal mode shapes can be found in Ref.~\onlinecite{dAquino:NormalModes}.

We have extended the method to be able to also compute the FMR spectrum of the simulated system, and report the new methodology in Appendix~\ref{subsec_appendix_eigenvalue} not to distract from the results obtained with the method.

\begin{figure}
  \includegraphics[width=1\columnwidth]{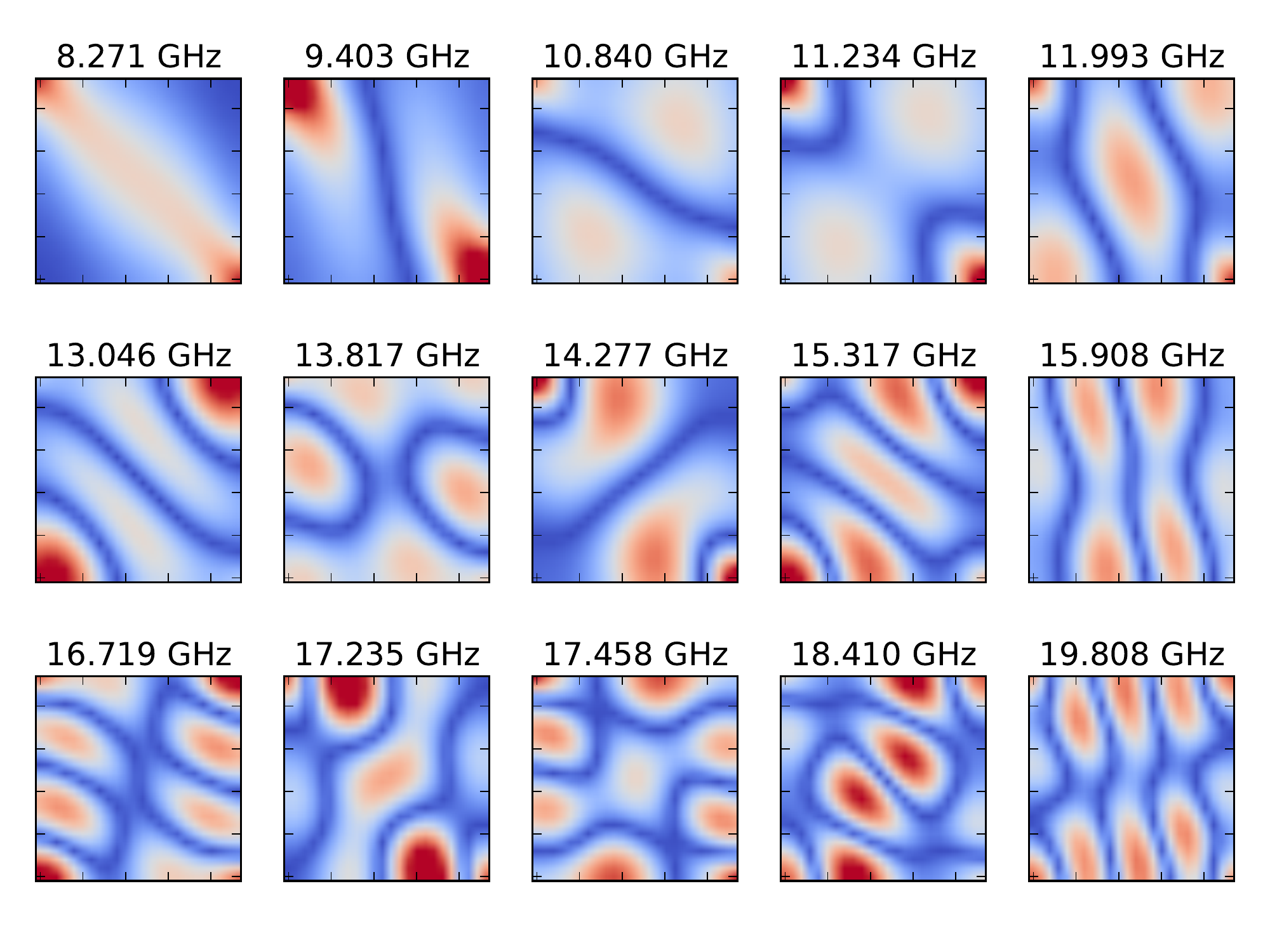}
  \caption{\label{Fig:fifteen_modes} The spatial power spectrum of $y$-component of magnetization for the 15 lowest frequency modes. The squares measure 120~nm of each side.}
\end{figure}

\begin{figure}
  \includegraphics[width=1\columnwidth]{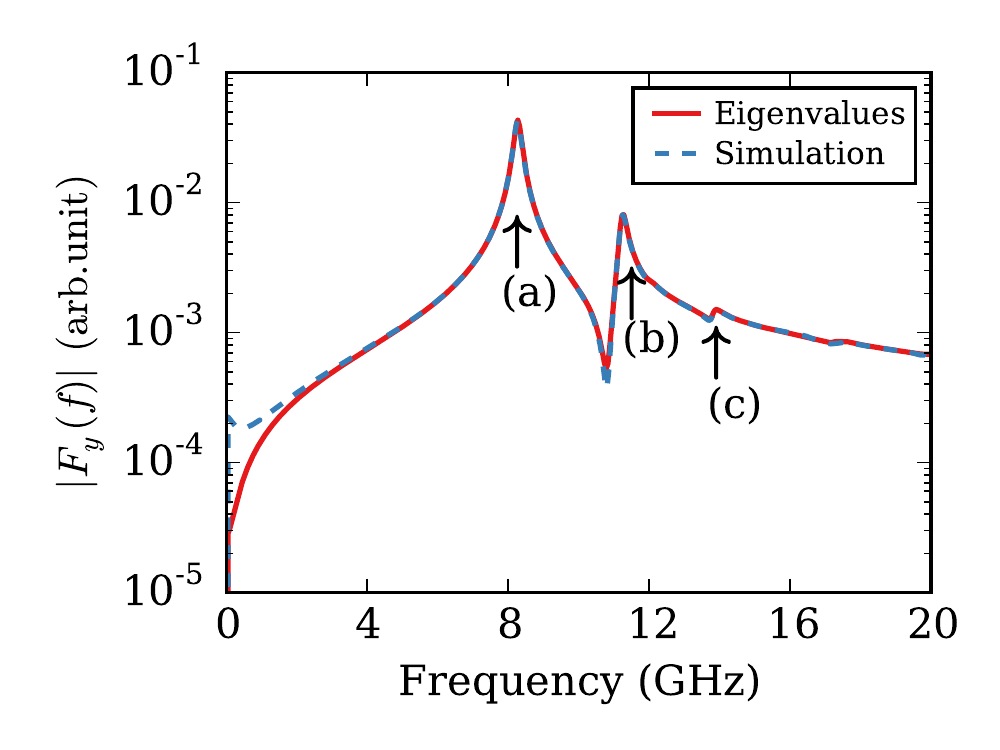}
  \caption{  \label{Fig:eigenvalueSpectrum}Comparison of the resonance spectra calculated using method 1 [$|F_{{y}}(f)|$, from Eq.\ (\ref{Eq:spatiallyAveragedFT})] obtained by simulation using \textsf{OOMMF} (dashed blue line) and from the eigenvalue problem formulation (solid red line). Excellent agreement is observed over the whole frequency range, although the peak heights are slightly different. Arrows denote the positions of the modes plotted in Fig.\ \ref{Fig:ringdownEigenComparison}.}
\end{figure}

Table~\ref{Tab:modeFreqs} shows the first fifteen resonance frequencies, calculated with the eigenvalue approach using a finite difference discretization with cell size $5 \times 5 \times 5 \,\text{nm}$, matching the simulation parameters used by \textsf{OOMMF}. The spatial distribution of these modes are plotted in Fig.~\ref{Fig:fifteen_modes}. The power density spectrum, and thus the amplitude of each mode excited during the simulation, is dependent upon the perturbation of the system. Using the method described in Appendix~\ref{subsec_appendix_eigenvalue_spectrum}, we compute the coupling of the used excitation to each mode and reconstruct the spectrum shown in Fig.~\ref{Fig:eigenvalueSpectrum}, demonstrating an excellent agreement between the ringdown method and the eigenvalue method.

\begin{figure}
  \includegraphics[width=1\columnwidth]{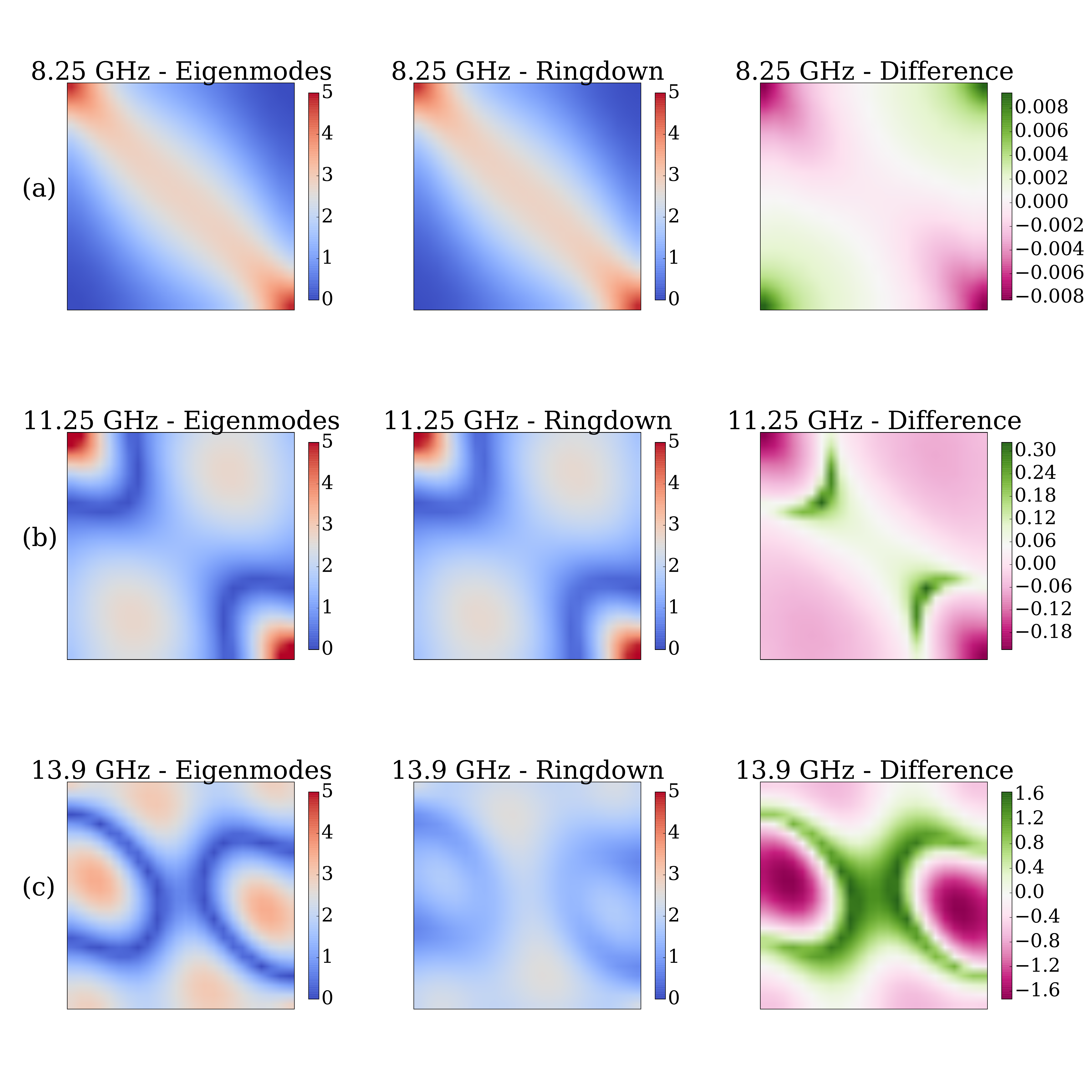}
  \caption{\label{Fig:ringdownEigenComparison} Comparison of the spatially resolved power spectrum given by the ringdown method 2 from \textsf{OOMMF} (middle column) and the eigenvalue problem (left column) for the $y$-component of the 3 lowest frequency modes (\textit{top row:} $8.25 \,\text{GHz}$, \textit{middle row:} $11.25 \,\text{GHz}$, \textit{bottom row:} $13.9 \,\text{GHz}$). Excellent agreement is observed for $8.25 \,\text{GHz}$ and $11.25 \,\text{GHz}$. The agreement gets worse as the amplitude of the mode generated by the ringdown method decreases, leading to a larger signal-to-noise ratio and a less well defined spatial plot.}
\end{figure}

Finally, we show the comparison of the spatial profiles generated by the ringdown and eigenvalue methods. Figure~\ref{Fig:ringdownEigenComparison} shows a comparison of the three lowest frequency modes, demonstrating excellent agreement for the two modes visible in Fig.~\ref{Fig:FTComparison}. This agreement gets worse as the frequency of the normal modes increases and their amplitude in the ringdown method decreases, increasing the signal-to-noise ratio; above $14 \,\text{GHz}$ the data quality is not sufficient to make a meaningful comparison. Nevertheless, the close agreement of results demonstrates the equivalence of these two approaches.

\begin{table}
  \begin{tabular}{c c}
    \hline
    \hline
    Mode Frequency (GHz) & Damping Time (ns) \\
    \hline
    8.270 & 1.549 \\
    9.402 & 1.639 \\
    10.839 & 1.437 \\
    11.233 & 1.452 \\
    11.992 & 1.401 \\
    13.045 & 1.345 \\
    13.816 & 1.292 \\
    14.276 & 1.253 \\
    15.316 & 1.191 \\
    15.907 & 1.156 \\
    16.718 & 1.126 \\
    17.234 & 1.094 \\
    17.457 & 1.094 \\
    18.409 & 1.030 \\
    19.806 & 0.963 \\
    \hline
    \hline
  \end{tabular}
  \caption{  \label{Tab:modeFreqs}Frequency and damping time of the 15 lowest frequency modes, calculated using an eigenvalue problem approach.}
\end{table}

%%%%%%%%%%%%%%%%%%%%%%%%%%%%%%%%%%%%
\subsection{Falsification Properties\label{subsec:Falsification}}

In defining a standard problem, it is useful to investigate how changing the parameters of the simulation will distort the results. This is intended to allow users to isolate inconsistencies within their own simulations when attempting to reproduce the output of this problem.

\subsubsection{Damping Parameter\label{subsubsec:damping}}

\begin{figure}
  \begin{center}
    \includegraphics[width=1\columnwidth]{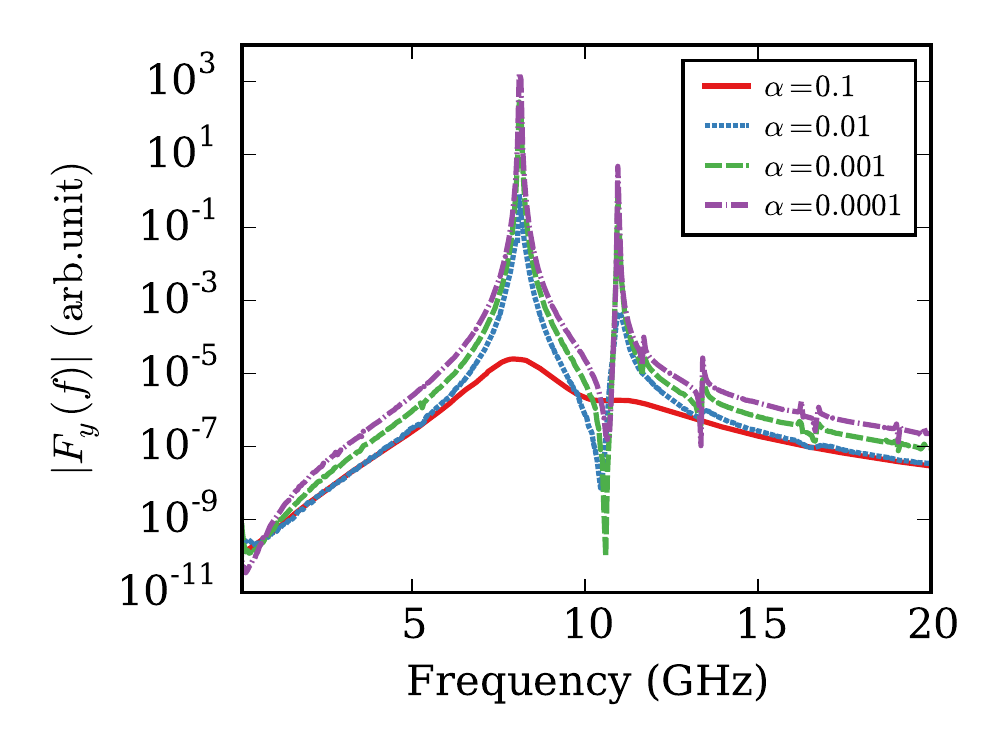}
  \end{center}
  \caption{\label{Fig:varyDamping_Spectrum} Normalized FMR spectrum for systems in the dynamic stage with a range of damping constants. At $\alpha \geq 0.5$ the system is over-damped, not producing resonance modes. As the damping decreases the peaks become taller and sharper.}
\end{figure}

The magnitude of the Gilbert damping parameter during ringdown method determines the time taken for the system to reach its stable configuration. However, this did not affect the resonance frequencies produced by the Fourier transform, except in the strongly damped case where $\alpha \geq 0.1$. Figure~\ref{Fig:varyDamping_Spectrum} shows the power spectrum produced by the simulation for $\alpha$ = $10^{-1}$, $10^{-2}$, $10^{-3}$ and $10^{-4}$. As the damping parameter is decreased, the peaks become narrower and taller as expected. For the highest damping the spectrum is heavily suppressed, showing only two broad features, with the $11.25 \,\text{GHz}$ mode barely visible above the tails of the $8.25 \,\text{GHz}$ mode. In this case the system is approaching overdamping; if we choose a large damping of $\alpha = 1$ then no precession occurs and the Fourier transform shows no peaks. As damping decreases extra peaks begin to form, for example at $f \approx$ 12~GHz and 13.5~GHz. At lower dampings the intensity of these features increases, but never surpasses that observed for the two dominant modes.

\subsubsection{Relaxation Time\label{subsubsec:relax}}
This standard problem definition (see Sec.~\ref{subsec:problemdef}) asks that the computation of the relaxed configuration should be carried out by integrating the damped equation of motion for 5ns.
In this subsection, we explore how the obtained frequency spectrum changes if a shorter period is used.

\begin{figure}
  \begin{center}
    \includegraphics[width=1\columnwidth]{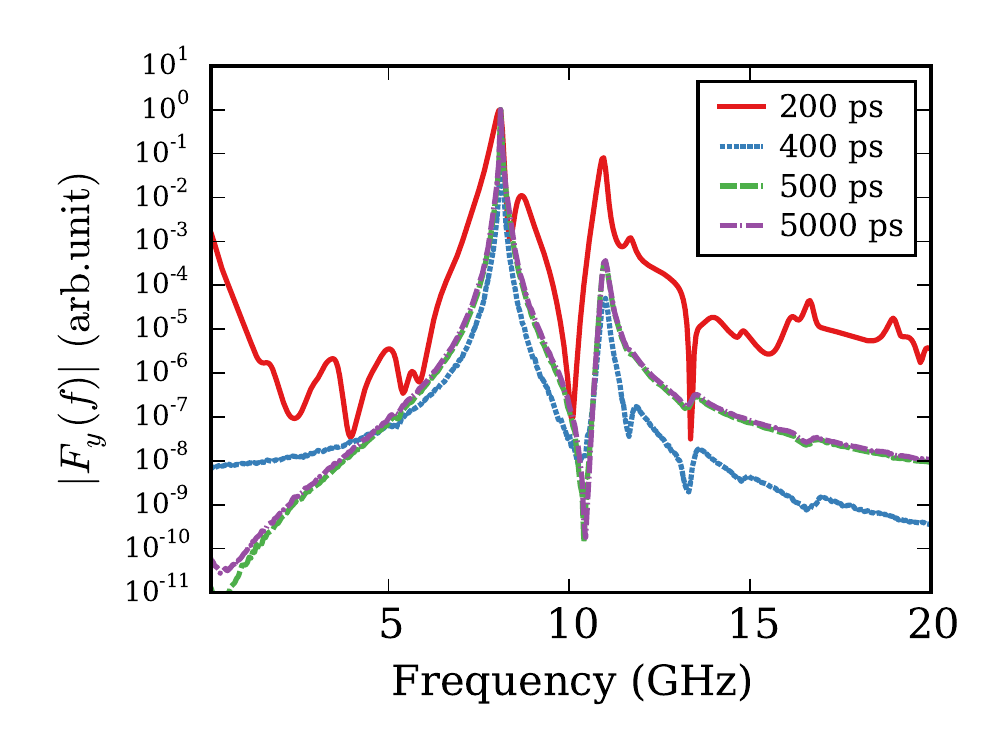}
  \end{center}
  \caption{\label{Fig:convExamples} Normalized FMR spectrum as calculated for systems entering the dynamic stage after varying the time in the relaxation stage. Allowing more time to relax leads to a lower amplitude, less noise and more well-defined peaks.}
\end{figure}

Figure~\ref{Fig:convExamples} shows that starting the dynamic simulation stage from an improperly converged configuration from the relaxation stage causes significant instability within the system. Although there are still peaks at the resonance frequencies for relaxation times shorter than 5ns, there are also many other peaks corresponding to domains aligned in other directions relaxing back to align with the bias field. The frequency of the normal modes does not change, but the strength of the contributions from spurious modes is too large to allow a meaningful analysis. The importance of allowing the relaxation stage sufficient time to reach a converged state is clear; the difference occurs because the the system dynamics contains the components that exist as a consequence of the system tending to reach the equilibrium state during the dynamic stage of simulation. We can see that the curves for $500 \,\text{ps}$ and $5000 \,\text{ps}$ produce very similar results.

\subsubsection{Relaxation Stage Perturbation Angle\label{subsubsec:excitation}}

\begin{figure}
  \includegraphics[width=8cm]{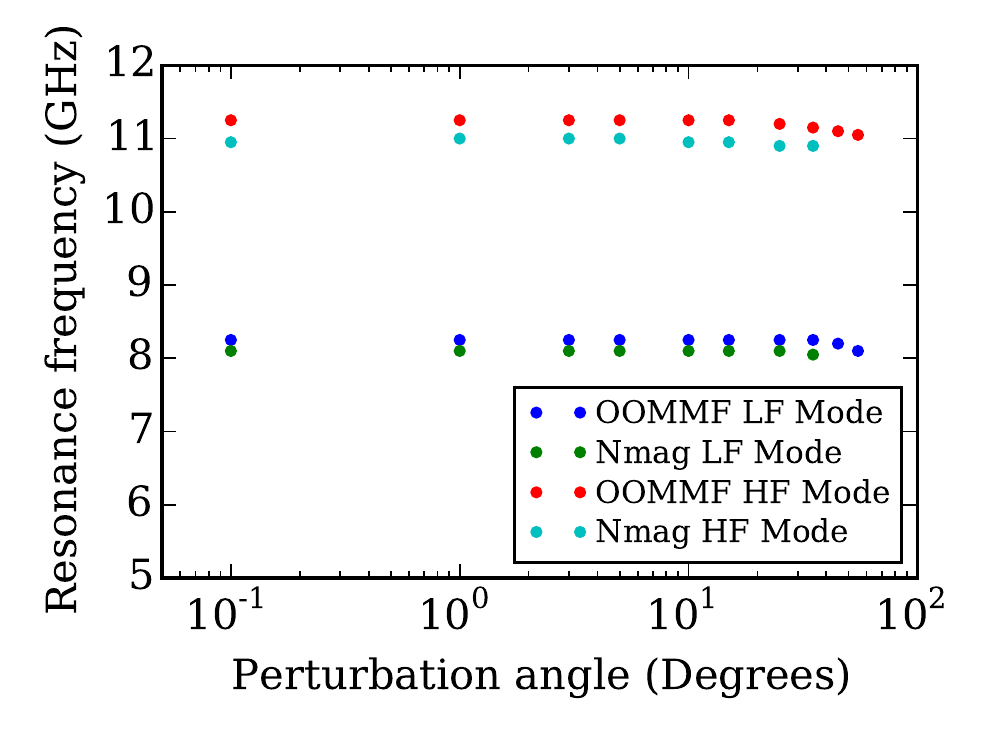}
  \caption{\label{Fig:vary_relaxAngle} Changes to the resonance frequency of the main modes in the FMR spectrum as the magnitude of the initial perturbation is altered. The FD method used by OOMMF is relatively unaffected by this change for small angles, at high angles the spectrum becomes noisy and resonance frequency drops. LF stands for Low Frequency peak at $\approx8\,$GHz and HF for High Frequency at $\approx11\,$GHz}
\end{figure}

The problem definition (Sec.~\ref{subsec:problemdef}) suggests a change of $0.56^\circ$ between the bias field in the relaxation and dynamic stage. Figure~\ref{Fig:vary_relaxAngle} shows the effect of changing the perturbation angle between bias field in the relaxation and dynamic stages from $0.1^\circ$ to $55^\circ$.

Changing the perturbation angle of the bias field changes the amount of energy supplied to the system in the initial excitation, which manifests as a greater area under the power spectrum curve (not shown). If the perturbation angle is too small ($< 0.1^{\circ}$), no peaks are observed above the noise level of the power spectrum.
Conversely, if the perturbation angle is too large the system deviates significantly from the equilibrium state, and additional modes form, leading to a distorted power spectrum. While the resonance frequency does not significantly change, the spectrum is eventually dominated by these other features and modes. At such high perturbations both \textsf{Nmag} and \textsf{OOMMF} show a slight drop in resonant frequency.
%HF: not sure what the next line tries to say:
%We note that there is a slight asymmetry in the effects of perturbations in the positive and negative $y$-directions, which we ascribe to the fact that a bias field angle of 35$^{\circ}$ is not symmetric in this direction.

\subsubsection{Spatial Discretization\label{subsubsec:mesh}}

In micromagnetics, it is generally recommended to keep the cell size smaller than the exchange length, and for this standard problem, we use a cell with an edge length of 5\,nm.

\begin{figure}
\includegraphics[width=8cm]{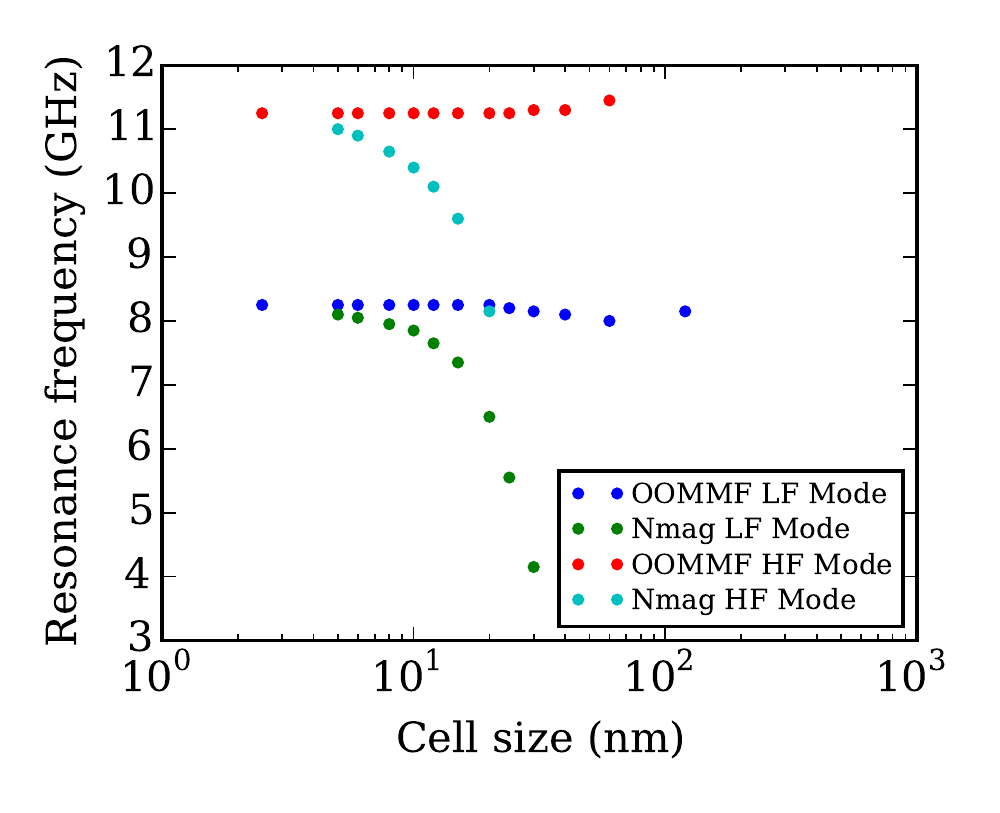}
\caption{Location of the main resonance modes in the FMR spectrum as a function of resolution of the mesh. The FE method shows greater deviation from standard results with changes to the parameters, due to its more sensitive handling of demagnetization effects.}
\label{Fig:vary_cellSzie}
\end{figure}

The effects of changing the cell size from $2.5\,$nm to $120\,$nm are shown in Fig.~\ref{Fig:vary_cellSzie}. Decreasing the resolution of the mesh (increasing the size of the tetrahedra in FE or cuboids in FD) causes the divergence between FD and FE codes. This is to be expected, as the differing approach to calculation of demagnetization field is one of the key differences between the two approaches. In \textsf{OOMMF} the frequency of the low frequency mode decreases, while the frequency of the main edge mode increases. This comes about due to changes in relative importance of demagnetization effects from the edge of the sample, with fewer nodes near the boundaries the sample becomes more like an idealized infinite thin film.

In both codes, high frequency features are suppressed with increasing element size. These correspond to higher order modes that cannot form if there are too few elements to support their spatial variation. It is well known that choice of an appropriate mesh discretization is crucial in computational micromagnetics, an aphorism that is well supported by these results. The deviation of resonant frequency with mesh resolution therefore suggests that a resolution comparable to the exchange length in permalloy ($\sim5 \,\text{nm}$) is appropriate. We can also see that \textsf{OOMMF}'s finite difference approach is more robust than \textsf{Nmag}'s finite element based result here.

%%%%%%%%%%%%%%%%%%%%%%%%%%%%%%%%%%%%
\subsection{Comparison of Simulation Methodologies\label{subsec:FeFd}}
\begin{figure}
\includegraphics[width=8cm]{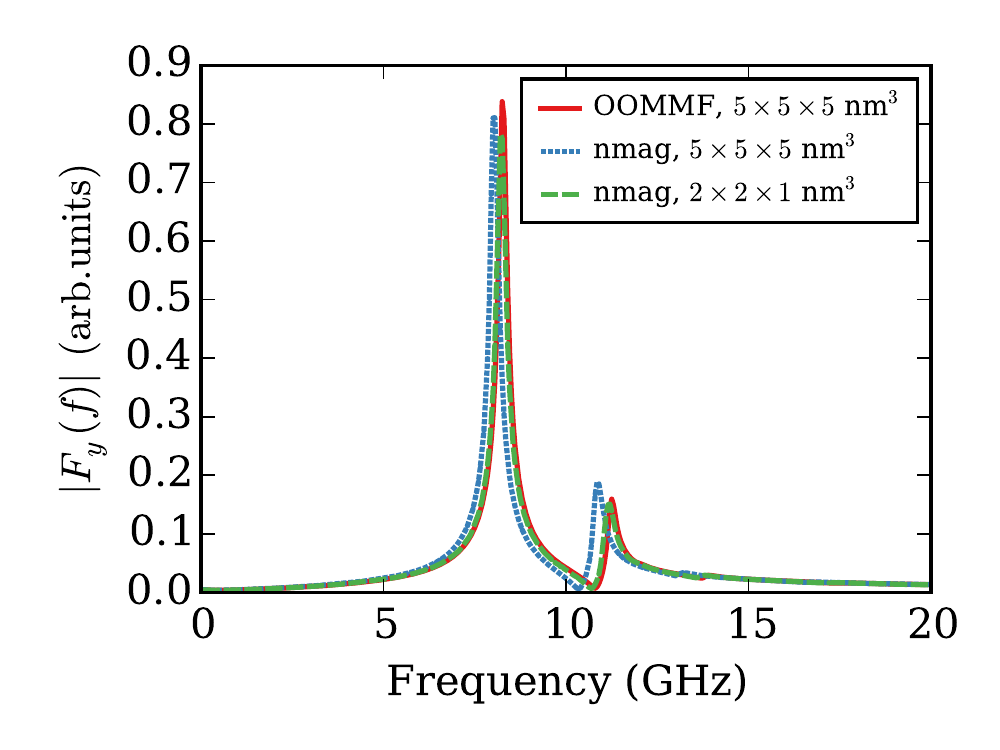}
\caption{Comparison of the resonance spectra $|F_y(f)|$ obtained using \textsf{OOMMF} and \textsf{Nmag} with different resolutions of the mesh.
A cell size of $5 \times 5 \times 5\,$nm$^{3}$ is used for \textsf{OOMMF}. Data from \textsf{OOMMF} and \textsf{Nmag} agree well when the cell size for \textsf{Nmag} is reduced.}
\label{fig_cmp_fy}
\end{figure}
For the standard problem defined above with a cell size of $5 \times 5 \times 5\,$nm$^{3}$, the deviations between finite difference and finite element methods for both the resonance frequencies are noticeable, as shown in Fig.~\ref{fig_cmp_fy}, reaching 0.2~GHz for the low frequency mode, and 0.4~GHz at the higher frequency. Note that in the case of the tetrahedra used in the finite element method this means that space is divided into 6 tetrahedra that together form a cube of dimensions $5 \times 5 \times 5 \,\text{nm}^{3}$.
A smaller cell size $2 \times 2 \times 1 \,\text{nm}^{3}$ for the finite element code will reduce the deviations significantly, bringing the two codes to within 0.05~GHz of agreement.

\begin{figure}
\includegraphics[width=7.6cm]{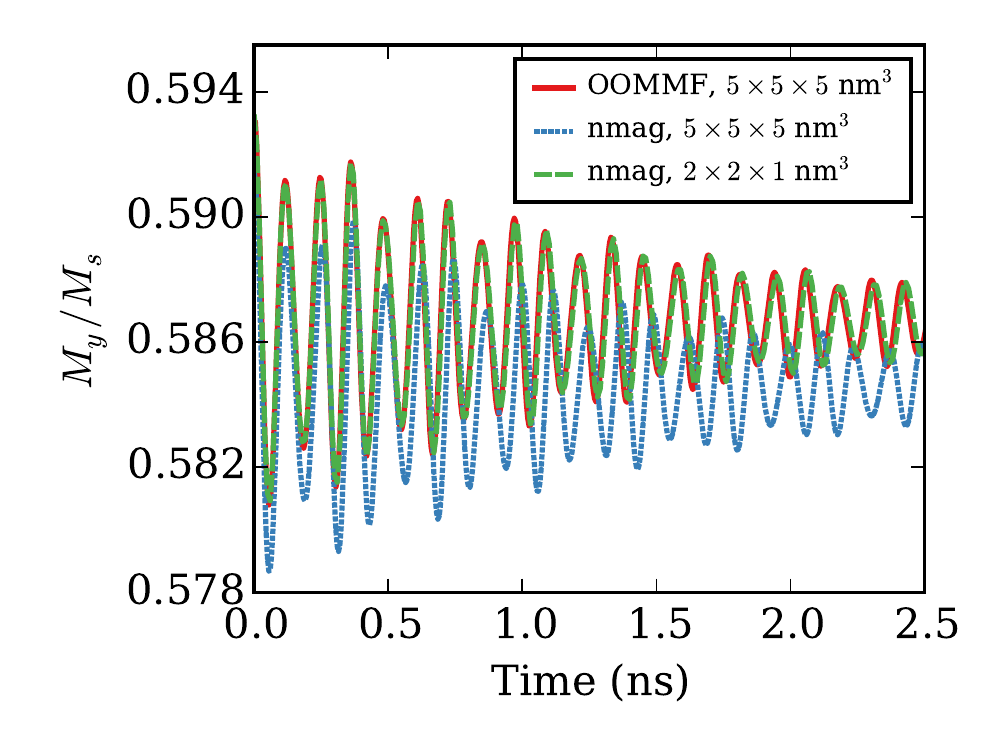}
\caption{Comparison of average magnetization ($y$-component) evolution between \textsf{OOMMF} and \textsf{Nmag} with different resolutions of the mesh. Note the phase shift that develops between different spatial resolutions in \textsf{Nmag}, corresponding to a different mode frequency in Fig.~\ref{fig_cmp_my}. Data from \textsf{OOMMF} and \textsf{Nmag} agree well when the cell size for \textsf{Nmag} is reduced. \label{fig_cmp_my}}
\end{figure}

 The corresponding comparison for the average magnetization ($y$-component) evolution is shown in Fig.~\ref{fig_cmp_my}. It is obvious that the two micromagnetic package produce different simulation results when the cell size is $4 \times 5 \times 5 \,\text{nm}^{3}$, but good agreement is found for the smaller cell size of $2 \times 2 \times 1 \,\text{nm}^{3}$.

%This behaviour can be understood by considering the different methods to compute the effective field based on the different approaches to spatial discretization employed by finite difference and finite element packages. Excellent agreement is obtained when a finer mesh discretization is used in \textsf{Nmag}.
%While finite difference and finite element methods produce the same value for the frequency of the thin film resonance peak there is a significant difference for the demagnetization mode. This deviation can be understood by considering its origin within the sample and the different approaches to spatial discretization employed by finite difference (FD) and finite element (FE) packages. The energy in the demagnetization mode is concentrated at the edges of the film [as shown in Fig.\ \ref{Fig:spatialModes}], and decreases sharply towards the middle.

%Computing the demagnetization field is a computationally expensive step using the FE method (FEM), converging much more slowly than when the FD method (FDM) is used.

In FD the computation takes place at the center of a series of cuboids used to build the sample, while in FEM it takes place at the nodes of the mesh tetrahedra. While tetrahedra give significantly better approximations to irregular shapes than the cuboids, computing values on vertices is problematic when the values of the demagnetization tensor vary sharply. If the mesh is not fine enough to accurately resolve the change, the effective fields will be calculated less accurately, and spurious results will be produced. In this simulation, the error arises from contributions from the top and bottom surfaces of the film, and a fourfold increase in resolution in the $z$-direction brings the FDM simulations into agreement with the FEM, at the cost of significantly increasing the runtime. This problem could also be alleviated through the use of a spatially varying mesh density, placing more mesh nodes in the regions near the surfaces to accurately sample the demagnetization tensor.

% (fangohr 11/11/2014) As part of the discussion, we may want to comment that there is deviation between FE and FD codes in other standard problems [if so]. Should check Problem 4 and 3, I think, maybe also the new problem number 5.

%%%%%%%%%%%%%%%%%%%%%%%%%%%%%%%%%%%%%%%%%%%%%%%%%%%%%%%%%%%%%%%%%
\section{Summary and conclusions}
\label{sec:summary}

A standard problem for micromagnetic simulations of ferromagnetic resonance in a thin film has been introduced. FMR is a technique that is widely used for material characterization and the study of spin transfer phenomena. While micromagnetic simulations are able to provide insightful analysis and prediction of FMR experiments, it is not trivial to conduct those simulations. With this paper, we provide step by step instructions and specific parameters and results that can be used to validate simulation tools before they are applied to new problems.

We provide performance data from two popular micromagnetics packages (\textsf{OOMMF} and \textsf{Nmag}), thus providing data for the deviations that can be expected between different discretization and computation strategies. This standard problem may serve as an introduction to the procedures involved and allow benchmarking and testing of new simulation packages.

Example scripts to run the simulations and analyse the data, as well as raw data for all the figures, are available in the associated electronic supplementary material.\cite{Githubrepo2015}

\section*{Acknowledgements}
A.A.B.\ acknowledges support from Diamond Light Source, and the EPSRC through a Doctoral Training Grant. This work has been supported through the EPSRC Centre for Doctoral Training grant EP/G03690X/1.

\appendix
%\section*{Appendix}

\section{\textsf{Nmag} tolerances}

By analyzing the time evolution of the average magnetization $z$-component, obtained by running the \textsf{Nmag} simulation with default time integration tolerances, we observe that numerical noise is present after approximatelly $0.8 \,\text{ns}$. By simply performing the Fourier transform on this data, this numerical noise can be interpreted as a particular eigenmode of certain frequency.  Although this does not affect any results presented in this work, we provide the following improved demagnetization field computation settings that suppress this:
\begin{lstlisting}
ksp_tols = {"DBC.rtol":1e-7,
             "DBC.atol":1e-7,
             "DBC.maxits":1000000,
             "NBC.rtol":1e-7,
             "NBC.atol":1e-7,
             "NBC.maxits":1000000,
             "PC.rtol":1e-3,
             "PC.atol":1e-6,
             "PC.maxits":1000000}
\end{lstlisting}
and time integration tolerances:
\begin{lstlisting}
sim.set_params(stopping_dm_dt=0.0, ts_abs_tol=1e-7, ts_rel_tol=1e-7)
\end{lstlisting}
in the dynamic simulation stage. The improved tolerances remove the numerical noise from the average magnetization time evolution, but increases the running time. The full scripts to run the simulations are available.\cite{Githubrepo2015}

\section{Eigenvalue approach\label{subsec_appendix_eigenvalue}}

In this Appendix, we provide a brief summary of the eigenvalue method described in Ref.~\onlinecite{dAquino:NormalModes}, with modifications required to compute the Gilbert damping and excitation dependent FMR spectrum of the system along with the resonance frequencies and corresponding normal modes.

The dynamics of the micromagnetic system is governed by the Landau-Lifshitz-Gilbert (LLG) equation:
\begin{equation}
  \label{Eq:normalModeLLG}
  \dot{\mvec{m}} = - \frac{\gamma}{1 + \alpha^{2}} \left[ \mvec{m} \times H_{\mathrm{eff}} + \alpha \mvec{m} \times \mvec{m} \times  H_{\mathrm{eff}} \right] = \mathcal{L}(\mvec{m}),
\end{equation}
where \mvec{m} is the normalized magnetization: $\mvec{m} = \mvec{M}/M_\text{s}$, with $|\mathbf{M}| = M_\text{s}$ being the saturation magnetization. If the system is in its equilibrium state $\mvec{m}_{0}$, then $\mathcal{L}(\mvec{m}_{0}) = 0$. Small perturbations from the equilibrium (for example, those generated by the removal of the external magnetic field perturbation when moving from the \textit{relaxation} to the \textit{dynamic} stage of simulation) can be described as $\mvec{m} = \mvec{m}_{0} + \varepsilon \mvec{v}$, with $\mvec{v} \perp \mvec{m}_0$ since the $|\mathbf{m}| = 1$ condition is imposed. For a small $\varepsilon$, terms of the $\mathcal{O}(\varepsilon^{2})$ order and higher can be neglected, which results in the linearized equation (for the general case):
\begin{equation}
  \label{Eq:normalModeLinearised}
  \dot{\mvec{v}} = \frac{\partial \mathcal{L}}{\partial \mvec{m}} \bigg|_{\mvec{m} = \mvec{m}_0} \mvec{v}.
\end{equation}
If the fixed linear operator $\hat{L} = \frac{\partial \mathcal{L}}{\partial \mvec{m}} \big|_{\mvec{m} = \mvec{m}_0}$ is defined, the linearized equation can be written as $\dot{\mvec{v}} = \hat{L}\mvec{v}$. This is an ordinary differential equation, which can be solved by an ansatz of the form $\mvec{v} = \mathrm{Re}(\tilde{\mvec{v}}e^{i2\pi ft})$. The normal modes (eigenvectors) $\tilde{\mvec{v}}$ and oscillation frequencies (eigenvalues) $f$ can be found from the following eigenvalue problem
\begin{equation}
  \label{Eq:normalModeEigenvalue}
  i2 \pi f \tilde{\mvec{v}} = \hat{L} \tilde{\mvec{v}}.
\end{equation}

\subsection{Linearized equation without damping\label{sec-linearized-nondamping}}

First, we consider the case when the damping term in the LLG equation is neglected ($\alpha=0$). Without damping, the magnetic moments precess indefinitely, and the LLG equation preserves energy. In this simplest case the calculation of the linearized operator $\hat{L}$ is fairly straightforward and results in the following linearized equation of motion~\cite{dAquino:NormalModes}
\begin{equation}
\dot{\mvec{v}} = \gamma \, \mvec{m}_0  \times \hat{A}\mvec{v}, \qquad \hat{A} = |H_\mathrm{eff}(\mvec{m}_0)| \operatorname{Id} -
\frac{\partial H_\mathrm{eff}}{\partial v}  ,
\label{Eq:linearized-nondamping}
\end{equation}
where~$\hat{A}$ is a positive definite Hermitian operator. The normal modes $\tilde{\mvec{v}}$ and frequencies $f$ of the linearized equation can be found from the eigenvalue problem
\begin{equation}
- i 2 \pi f \mvec{m}_0 \times \tilde{\mvec{v}}  = \gamma \, \hat{A} \tilde{\mvec{v}}.
\label{Eq:linearized-nondamping-problem}
\end{equation}
The left-hand side of this eigenvalue problem also contains a Hermitian operator describing the uniform precession $\hat{B}v = - i \, \mvec{m}_0 \times \tilde{\mvec{v}}$, however it is not positive definite (its eigenvalues are $\pm 1$).

Because of the energy conservation, the oscillation frequencies~$f_{k}$ that satisfy this eigenvalue problem will be real and the normal modes $\tilde{\mvec{v}}_k$ corresponding to different frequencies will be orthogonal. These properties enable the efficient numerical solution of Eq.\ \eqref{Eq:linearized-nondamping-problem}; the eigenvalues~$f$ are the resonant frequencies and the complex magnitudes of the eigenvectors~$\tilde{\mathbf{v}}$ are the normal mode amplitude plots (the complex phase of $\tilde{\mathbf{v}}$ corresponds to the phase of the oscillations at the corresponding sites).

However, in order to compute the FMR spectrum via the eigenvalue approach, we have to consider the more complicated case of non-zero damping.

\subsection{Linearized equation with damping --- perturbative analysis}

For the case of sufficiently small non-zero damping~$\alpha$, a perturbative analysis can be performed to determine the corrections to the eigenvalues. In this case, eigenvalues have the form $\lambda = i 2\pi f - 1 / \tau$, where $\tau$ is the characteristic time for the mode to decay to $1/e$ of its starting amplitude value. It turns out that to the first order, the resonance frequencies are unchanged, and the damping times can be found using a relatively simple analytic calculation without solving the perturbed eigenvalue equation numerically.\cite{dAquino:NormalModes} Additionally, the coupling between the perturbed normal modes is small if their frequencies are sufficiently separated --- this property will be useful for the calculation of the FMR spectrum. We have found that for our test system, which has a low damping constant, the damping times computed using both the perturbative method and the numerical method (from the next section) are very close; up to the 4 digits shown in Table~\ref{Fig:fifteen_modes} the results are identical for both methods.

\subsection{Linearized equation with damping --- numerical solution}

To compute the actual FMR spectrum, we have to derive the linearized equation in the presence of damping, and solve the corresponding eigenvalue problem. The derivation of the linearized equation with damping is straightforward but slightly tedious. We skip this derivation and instead compute the linearized equation using a numerical differentiation trick.

For the linearized equation, we have to compute the directional derivative
\begin{equation}
\hat{L} v = \bigg(\frac{\partial \mathcal{L}}{\partial \mvec{m}} \bigg|_{\mvec{m} = \mvec{m}_0}\bigg) [ v] = \frac{d}{d \epsilon} \mathcal{L}(m_0 + \epsilon v) |_{\epsilon = 0}.
\end{equation}
For the test problem, the components of the effective field (demagnetization, exchange, bias) are all either constant, or linear functions of $\mvec{m}$; therefore as a function of~$\epsilon$, $\mathcal{L}(m_0 + \epsilon v)$ is a degree 3 polynomial (the highest degree coming from the damping term $\mvec{m} \times \mvec{m} \times H_\mathrm{eff}$). This means that a numerical differentiation rule of order 3 or higher will compute the derivative $\frac{d}{d \epsilon} \mathcal{L}(m_0 + \epsilon v) |_{\epsilon = 0}$ exactly.

\subsection{Linearized equation with damping --- spectrum computation}
\label{subsec_appendix_eigenvalue_spectrum}

The previous sections outlined the method used to compute the frequencies and normal mode shapes for the linearized equation, with or without damping. In this section we describe the subsequent computation of the FMR spectrum, which also depends on the initial state of the system. To determine the contributions of each normal mode to the total spectrum, we have to compute the coupling between the initial state and the normal modes in the presence of damping. More precisely, let $n$ be the total number of the degrees of freedom (for a mesh with $N$ nodes, $n = 3N$). Let $\tilde{v}_{i}$, $i = 1, 2, \dots, n$ be the set of eigenvalues (normal modes) without damping, $\tilde{v}_i^{(p)}$, $i = 1, 2, \dots, n$ the set of perturbed eigenvalues in the presence of damping, and $f_i$ and $\tau_i$ the corresponding mode frequencies and damping times. Let $v_\mathrm{initial} = m_\mathrm{initial} - m_0$ be the initial state of the system. Due to the orthogonality property described in Sec.~\ref{sec-linearized-nondamping}, we can assume that with the respect to the Hermitian inner product defined by the operator $\hat{A}$ via $(x, y) := x \cdot \hat{A} \cdot y^*$ the non-perturbed eigenvectors form an orthonormal basis, i.e. $(v_i, v_j) = \delta_{ij}$.

To solve the linearized equation of motion Eq.~\eqref{Eq:normalModeLinearised}, we need to expand the initial state $v_\mathrm{initial}$ in the perturbed $\tilde{v}_i^{(p)}$ basis:
\begin{equation}
  v_\mathrm{initial} = \sum_{i=1}^n C_i \tilde{v}_i^{(p)}.
\end{equation}

Once the coefficients $C_i$ are known, the full solution of the linearized equation~\eqref{Eq:normalModeLinearised} is
\begin{equation}
  \label{eq-fullsolution-fmr}
  \mvec{m}(t) = \mvec{m_0} + \sum_{i=1}^{n} C_i e^{(2 \pi i \omega_i - 1 / \tau_i) t}.
\end{equation}
Given this full analytic solution, we can then calculate the spectrum using either of the methods described in Sec.~\ref{subsec:data}. Unfortunately, this expansion requires the knowledge of the complete set of eigenvectors~$\tilde{v}_i^{(p)}$, which is numerically unfeasible to compute. Instead, we will attempt to reconstruct the spectrum based on the first~$k$ perturbed modes with the lowest frequencies (we used $k=40$). We would like to do this by finding the ``best'' approximation
\begin{equation}
  v_\mathrm{initial} = \sum_{i=1}^k c_i \tilde{v}_i^{(p)} + R.
\end{equation}
We will look for this approximation in the subspace spanned by the first $k'$ non-perturbed normal modes $\tilde{v}_{i}$, with $k' > k$ (we used $k'=60$). Due to the frequency separation property mentioned earlier, we can expect that this restriction will not affect the residue (the high-frequency modes will not measurably contribute to the low-frequency spectrum). When restricted to this subspace, we arrive at a system of $k'$ equations with $k$ unknowns $c_i$
\begin{equation}
  (v_\mathrm{initial}, v_j) = \sum_{i=1}^k c_i (\tilde{v}_i^{(p)}, v_j), \quad i=1\dots k'.
\end{equation}
This linear system is overspecified but any residue will only contain the high-frequency modes with frequencies above $f_k'$, which don't contribute to the spectrum for frequencies below $f_k$ that we are trying to compute. We solve this linear system with a standard linear least-squares (linear regression) method, allowing the determination of the coupling coefficients~$c_i$ and thus the full solution Eq.\ \eqref{eq-fullsolution-fmr}.

\section{Simulations without Demagnetization}
\label{subsec_appendix_noDemag}
\begin{figure}
\includegraphics[width=1\columnwidth]{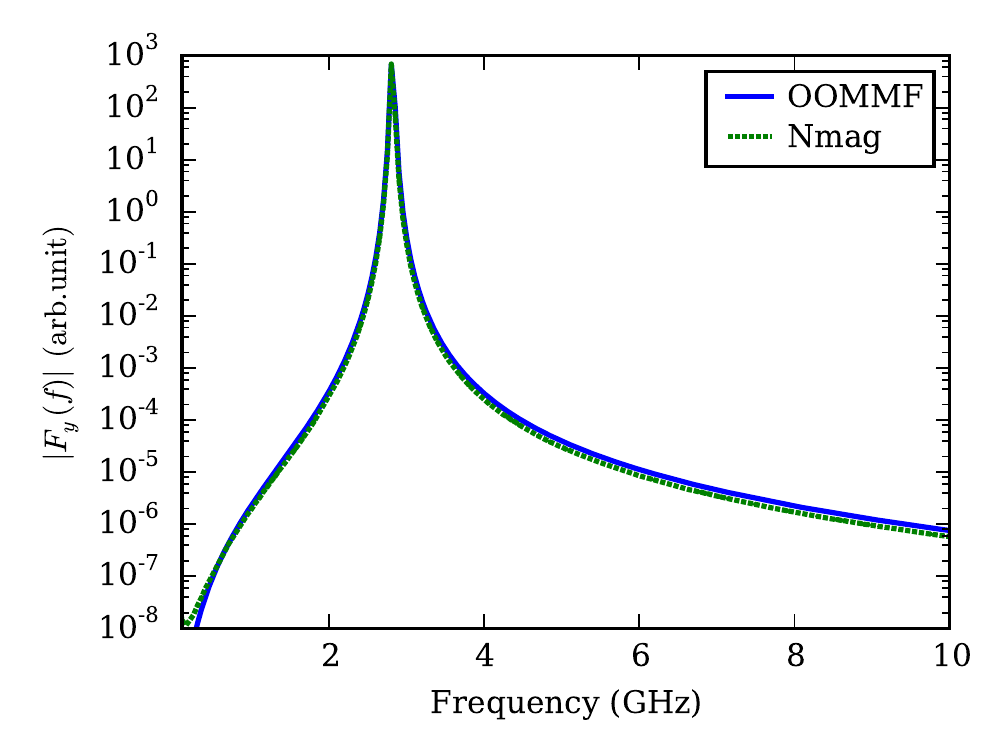}
\caption{Power spectrum for the proposed standard problem with the demagnetization field disabled. This removes all but one peak, wherein the entire sample is in resonance together. Finite element and finite different codes produce the same result under these conditions.}
\label{Fig:noDemag}
\end{figure}

In this section, we show the results of this standard problem in a
setup where only the exchange and the applied Zeeman effective fields
are considered. In particular, the demagnetization energy has been
ignored.
%These results are therefore directly comparable with those
%presented in Fig. \ref{fig_cmp_fy}.
Figure~\ref{Fig:noDemag} shows the
power spectrum of a simulation carried out with demagnetization
effects disabled in \textsf{OOMMF} and \textsf{Nmag}. The data has
been obtained using the ringdown method. It can be seen that the two
packages are in excellent agreement, producing only one mode at $2.8
\,\text{GHz}$. In the absence of the demagnetization energy, we obtain
this single mode corresponding to coherent precession of the
magnetization as shown in Fig.~\ref{Fig:noDemag}. This matches the
result from the Kittel equation for a material when demagnetization
energy contribution is neglected, for which:\cite{kittel_textbook}
\begin{equation}
  f = \frac{\gamma}{2\pi} \times \mu_{0} \times H_{\mathrm{applied}},
\label{Eq:noDemagKittel}
\end{equation}
yields $f = 2.81 \,\text{GHz}$.

As the simulation starts from a uniform, well-converged state only the lowest order, uniform, mode is observed. Modes located at the edge of the sample are suppressed due to the absence of demagnetization.

As discussed in Sec.~\ref{subsec:FeFd} differences can arise between simulations performed using the finite difference and finite element approaches due to their handling of demagnetization effects at the film boundaries. The data above shows that both approaches produce very similar spectra in the absence of demagnetization effects. We stress that this information is presented for comparative purposes only - it does not have physical meaning. Running simulations without demagnetization is, however, a useful tool in the debugging process or to analyze specific effects without the additional complications of magnetostatic energy.

\section{Software packages used}
\label{subsec_softwareDetails}

\begin{itemize}
  \item \textsf{OOMMF} version: 1.2 alpha 6
  \item \textsf{Nmag} version: 0.2.1
  \item \textsf{Python} version: 2.7.8 or 3.5.1
  \item \textsf{Numpy} version: 1.10.4
  \item \textsf{Scipy} version: 0.17.0
  \item \textsf{Matplotlib} version: 1.5.1
\end{itemize}

%\bibliography{FMR_StandardProblem_References}

%% ***** In final version for journal submission:
% Comment out \bibliography{FMR_StandardProblem_References}
% Paste here the bbl files
%% *****


\end{document}